\documentclass{aastex6}

\fullcollaborationName{The Friends of AASTeX Collaboration}

\begin{document}

%% LaTeX will automatically break titles if they run longer than
%% one line. However, you may use \\ to force a line break if
%% you desire.

\title{Dark Matter-Dark Energy Interaction and the Shape of Cosmic Voids}

%% Use \author, \affil, plus the \and command to format author and affiliation
%% information.  If done correctly the peer review system will be able to
%% automatically put the author and affiliation information from the manuscript
%% and save the corresponding author the trouble of entering it by hand.
%%
%% The \affil should be used to document primary affiliations and the
%% \altaffil should be used for secondary affiliations, titles, or email.
%% Authors with the same affiliation can be grouped in a single
%% \author and \affil call.
\author{Zeinab Rezaei}
\affil{Department of Physics, {School of Science}, Shiraz
University, Shiraz 71454, Iran.\\
Biruni Observatory, {School of Science}, Shiraz
University, Shiraz 71454, Iran.}

%% Notice that each of these authors has alternate affiliations, which
%% are identified by the \altaffilmark after each name.  Specify alternate
%% affiliation information with \altaffiltext, with one command per each
%% affiliation.

%% Mark off the abstract in the ``abstract'' environment.
\begin{abstract}
Interaction between dark matter and dark energy as one of not completely
solved problems in cosmology has been studied extensively. This interaction
affects the cosmic structures. In this regard, the shape of cosmic voids can be
influenced by the dark matter and dark energy interaction. Here, employing the dynamical dark energy model constrained by the observational data, we study the effects of this interaction on the ellipticity of cosmic voids. To this aim, we apply the linear growth of density perturbation in the presence of interaction.
The probability density distribution for the ellipticity of cosmic voids has been investigated.
The results confirm that the ellipticity of cosmic voids increases when the dark matter
and dark energy interaction is considered.
\end{abstract}

%% Keywords should appear after the \end{abstract} command.
%% See the online documentation for the full list of available subject
%% keywords and the rules for their use.
\keywords{Dark matter, Dark energy, Large-scale structure of Universe.}

%% From the front matter, we move on to the body of the paper.
%% Sections are demarcated by \section and \subsection, respectively.
%% Observe the use of the LaTeX \label
%% command after the \subsection to give a symbolic KEY to the
%% subsection for cross-referencing in a \ref command.
%% You can use LaTeX's \ref and \label commands to keep track of
%% cross-references to sections, equations, tables, and figures.
%% That way, if you change the order of any elements, LaTeX will
%% automatically renumber them.

%% We recommend that authors also use the natbib \citep
%% and \citet commands to identify citations.  The citations are
%% tied to the reference list via symbolic KEYs. The KEY corresponds
%% to the KEY in the \bibitem in the reference list below.

\section{Introduction} \label{sec:intro}

{Based on the observational data, the dark matter (DM) and dark energy (DE) interact non-gravitationally \citep{1311.7380,1603.08299}.
SNIa Union $2.1$ data have been applied to reconstruct the interaction between DE and DM \citep{1505.04443}.
}%
{Cosmological models with direct couplings between DM and DE have been probed by the cosmological observations \citep{vandebruck2018}.
}%
{Coupling between DE and DM with an energy transfer either from DM to DE or the opposite has been constrained by the cosmological data} \citep{
Guo(2007),1004.4562,1012.3904,1212.2541,1401.5177,1401.2656,Valiviita(2015),1605.01644,Murgia(2016),1705.09278,1711.06799,vandebruck2018,yang2018}.
{Different observational data such as cosmic microwave background (CMB) shift parameter, baryon acoustic oscillation (BAO), lookback time and Gold supernovae sample \citep{0902.0318}, optical, X-ray and weak lensing data from galaxy clusters \citep{0710.1198}, and relaxed galaxy clusters \citep{0910.5236} show energy decay from DE into DM \citep{1812.03540}. The energy transfer from DM to DE has also been suggested \citep{1403.4318,kumar2017}.
}

{The effects of the interaction coupling between dark components on the evolution of the Universe have been studied \citep{0901.3272,Zimdahl(2001),Binder(2006),1009.1214,Valiviita (2008)}. Coupling between DM and DE is compatible with an accelerated expansion of the Universe} \citep{Zimdahl(2001),0901.3272}.
{A non-relativistic DM component which interacts with DE can affect the behavior of the deceleration parameter, density parameters, and luminosity distance} \citep{Binder(2006)}.
{Interaction between DE and DM changes from negative to positive as the expansion of our Universe changes from decelerated to accelerated one }\citep{Valiviita (2008),1009.1214}.
{Solving the coincidence problem considering the coupling between DE and DM has been explored \citep{0707.2089,0801.4233}.
}

{Interaction between DE and DM has been studied using the CMB data \citep{Pavon(2004),0610806,1711.05196}.
Employing the WMAP results for the location of the CMB peaks shows that interacting models are consistent with the observational bounds \citep{Pavon(2004)}.
}%
{Presence of a constant coupling between DE and DM increases the tension between the CMB data from the analysis of the shift parameter in models with constant DE equation of state parameter and SNIa data \citep{0610806}.
}%
{Effects of interaction on the CMB and linear matter power spectrum have been investigated \citep{1711.05196}.
}

{DM-DE interaction can affect the structure formation at different scales and times.
}%
{Coupling between DE and cold DM (CDM) fluid results in the evolution of CDM and baryon distributions in structure formation \citep{0812.3901}, instability in the dark sector perturbations at early times \citep{Valiviita (2008)}, large-scale deviations in the power spectrum \citep{1502.06424}, and late-time transitions from DM to DE domination \citep{1707.09246}.
}%
{Structure growth has been investigated in the coupled DE models \citep{Caldera-CabralG(2009),Simpson(2010),1007.3736,1305.0829,web,1407.7548,1503.07875,1504.07397,2015Giocoli,1604.04222,1703.07357}.
An enhancement (suppression) of the growth factor takes place when the energy transfer in the background is from DM to DE (DE to DM) }\citep{Caldera-CabralG(2009)}.
{The growth of large-scale structure is suppressed because of the elastic interaction between DM and DE \citep{Simpson(2010)}.
}%
{Standard coupled DE model gives the enhanced growth of linear CDM density fluctuations \citep{1305.0829}.
}%
{DM and DE coupling increases the growth of the perturbations and the effective friction term in non-linear dynamics \citep{1407.7548}.
}%
{Coupling in the dark sector which leads to a bulk dissipative pressure can suppress structure formation at small scales \citep{1504.07397}.
Considering the DM-DE interaction, structure growth is suppressed and the tension between CMB observations and structure growth inferred from cluster counts can be explained \citep{1604.04222}.
}%
{The structure growth rate is also influenced by the DM-DE interaction \citep{0803.2239,0709.1128,1605.05264,1707.07667}.
}%
{With the DM-DE couplings, the structure growth rate shows a different time evolution \citep{0803.2239}.
}%
{Considering the large coupling strength, the instability leads to the exponential growth of small scale modes \citep{0709.1128}.
}%
{Employing the GLAMER gravitational lensing code verifies that
the coupling between DE and CDM is confirmed from the redshift evolution of the normalization
of the convergence power spectrum as well as the non-linear structure formation \citep{2015Giocoli}.
Applying the growth index parametrization, an analytic formula for the growth rate of structures in a coupled DE model has been presented }\citep{1605.05264}.
{Disformal coupling between DM and DE results in the intermediate scales and time dependent damped oscillatory features in the matter growth rate function \citep{1707.07667}.
}

Cosmic voids, the empty spaces in the large-scale structure of the Universe, are influenced by DE. Dynamical DE component forms these voids in response to the gravitationally collapsing matter \citep{0612027}.
Clustering of DE and the DE density perturbations alter the structure formation within voids \citep{0709.2227}.
The evolution of the ellipticity of cosmic voids is an important probe of the DE equation of state \citep{1-0704.088,5-1106.1611,1503.07690}.
The properties of DE affect the shape of voids through the ellipticity distribution of voids in large-scale structures \citep{1002.0014}.
Cosmological simulations employing different models for DE verify the sensitivity of void shapes to the nature of DE \citep{426-1-440}.

Interaction between DM and DE also alters the cosmic voids \citep{0903.0574,1406.0511,eliyv,11,1703.04885,1807.02938,2018hashim}.
In cosmological models with coupling between DE and DM, a further contribution to Integrated Sachs Wolfe effect arises during matter dominated era due to the DE perturbations associated with very large voids of matter \citep{0903.0574}.
Void catalogs can be applied to distinguish a model of coupled DM-DE from $\Lambda CDM$ cosmology due to the properties of cosmic voids \citep{1406.0511}. DM-DE coupling leads to larger voids as well as broader, shallower, and undercompensated profiles for large voids \citep{1406.0511}.
{Applying two void finders and a halo catalogue
extracted from the CODECS simulations which are the largest suite of publicly available cosmological
and hydrodynamical simulations of interacting DE cosmologies,
the interacting DE models have been investigated \citep{eliyv}.}
Filling factor, size distribution, and stacked profiles of cosmic voids can be affected by the DE coupling \citep{11}.
In cosmological models with evolving and interacting dark sectors, N-body simulations confirm that the presence of a coupled dark sector can be observable through the void statistics \citep{1703.04885}. The coupling of the dark sector slows down the evacuation of matter from voids \citep{1703.04885}.
Studying the structural properties of cosmic voids in a coupled DM-DE model confirms that the void merger rate within this model is greater than in $\Lambda CDM$ one \citep{1807.02938}. Coupled DM-DE model results in more large voids and delays the matter evacuation from voids because of the drag force acting on baryonic particles moving out of voids \citep{1807.02938}.
{N-body simulations of interacting dark energy model have been done to explore the structural properties of the cosmic voids \citep{2018hashim}.
Their results show that the internal structural properties of cosmic voids such as void density profile are different
from the $\Lambda CDM$ one.}
According to the above discussions, we can deduce that the DM-DE interaction can affect the shape of cosmic voids. Here, we study
the ellipticity of cosmic voids in the evolving dark energy models considering the DM-DE interaction.

\section{{Cosmological model with dynamical interacting dark energy} }

{In this section, we describe the cosmological framework in which we investigate the cosmic voids with interacting DE. First, the energy exchange between DM and DE and the approach to quantify the coupling between dark sectors are presented. Then, applying the dynamical DE equation of state, we illustrate the linear growth of density perturbation in the coupled cosmology.
}
\subsection{Interacting dark sectors}
To describe the DM-DE interaction, we introduce $\rho^{\rm int}$ as the DM-DE interaction energy density.
$\rho^{\rm int}$ denotes the energy density transfer from DE to DM. The rate of energy density exchange is also given by \citep{1705.09278,0801.4233,Valiviita (2008),0901.3272,1403.4318,1608.02454,0411025,0610806},
\begin{eqnarray}
Q=\frac{d\rho^{\rm int}}{dt},
 \end{eqnarray}
in which $Q>0$ means that the direction of energy transfer is from DE to DM and for $Q<0$ this direction is from DM to DE.
The rate of energy density exchange is proportional to the DM energy density, i.e. $Q=\eta H \rho_{\rm DM}$ \citep{1705.09278,0801.4233,1403.4318,1608.02454,0411025,0610806}.
The parameter $\eta$ determines the coupling between dark sectors or strength of the interaction between DM and DE.
This parameter characterizes the degree of the deviation from the noninteracting DE and DM. The value $\eta=0$ presents the noninteracting case. In addition, $H$ is the Hubble parameter.

One of the important attempts in studying the DM-DE interaction is constraining the value of $\eta$ using the observational data. In the present work, we apply the constrained value of the coupling parameter in the interacting dark
energy model using the observational data \citep{1705.09278} including the cosmic chronometer data, latest estimation of the local Hubble parameter value, joint light curves sample, BAO distance measurement data set, and the CMB data from Planck 2015 measurements.

\subsection{Linear growth of density perturbation}

We start from the Friedmann equations (applying the units with c = 1)
\begin{eqnarray}\label{f1}
\dot{a}^2+k =\frac{8\pi G}{3}\rho a^2,
 \end{eqnarray}
\begin{eqnarray}\label{f2}
\dot{a}^2+k +2a \ddot{a} =-8\pi G P a^2.
 \end{eqnarray}
Here, the total energy density, $\rho$, is given by the matter
energy density, $\rho_{\rm M}$, radiation energy density, $\rho_{\rm R}$, and DE density, $\rho_{\rm DE}$, i.e. $\rho=\rho_{\rm M}+\rho_{\rm R}+\rho_{\rm DE}$. Moreover, the total pressure, $P$,
is related to the matter pressure, $P_{\rm M}$, radiation pressure, $P_{\rm R}$, and DE pressure, $P_{\rm DE}$, by $P=P_{\rm M}+P_{\rm R}+P_{\rm DE}$. In addition, $k=-1,0$, and $1$ for an open, flat, and closed universe, respectively.
In the present work, we neglect the matter pressure, i.e. $P_{\rm M}=0$, and apply the radiation EOS $P_{\rm R}=\frac{1}{3}\rho_{\rm R}$. Besides, the DE EOS is considered as $P_{\rm DE}= w_{\rm DE}(z) \rho_{\rm DE}$, with two
different DE parameterizations as follows,
\begin{eqnarray}\label{DE1}
w_{\rm DE1}(z)=-1.0,
 \end{eqnarray}
\begin{eqnarray}\label{DE3}
w_{\rm DE2}(z)=w_{0}-w_{\beta}[\frac{(1+z)^{-\beta}-1}{\beta}].
 \end{eqnarray}
In above equations, the redshift $z$ is related to the scale factor by $a=(1+z)^{-1}$.
The first model is $\Lambda$CDM and the second one is a generalized evolving equation of state \citep{Barbozaetal}. Here, we consider the values of $w_{0}$, $w_{\beta}$, and $\beta$ obtained in the interacting dark energy model \citep{1705.09278} constrained using the observational data (i. e. $w_{0}=-0.944$, $w_{\beta}=-0.419$, and $\beta=0.350$).
It should be noted that in the following calculations we apply the $w_{\rm DE1}$ for DE EOS of noninteracting dark energy model and  the $w_{\rm DE2}$ for DE EOS of interacting one.
\begin{figure*}
\vspace*{1cm}       % Give the correct figure height in cm
\includegraphics[width=0.30\textwidth]{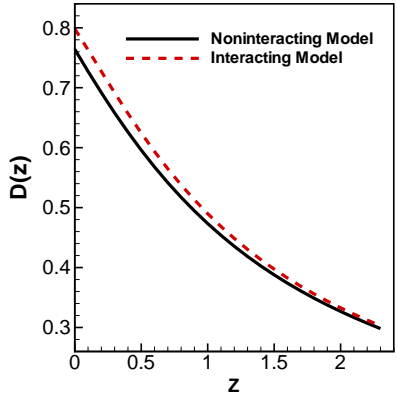}
\caption{Linear growth factor, $D$, versus the redshift, $z$, in noninteracting and interacting models.}
\label{fig5}
\end{figure*}

The linear growth of the matter density perturbation, $D(a)=\delta\rho_{\rm M}/\rho_{\rm M}$, is as follows  \citep{Bonnor,Heath,Percival,1-0704.088},
\begin{eqnarray}\label{Heath1}
D(a)=\frac{5\Omega_{\rm M}}{2}E(a)\int_0^a\frac{da'}{[a'E(a')]^3}.
 \end{eqnarray}
In the above equation, $E(a)\equiv H/H_0$. In a spatially flat universe in which $k = 0$ and for the case of noninteracting DM and DE, the function $E(a)$ is given by \citep{Percival,1-0704.088},
\begin{eqnarray}\label{Heath1NI}
E_{\rm NI}(a)=[\Omega_{\rm R} a^{-4}+\Omega_{\rm M} a^{-3}+\Omega_{\rm DE}\ a^{f(a)}]^{1/2},
 \end{eqnarray}
where $\Omega_{\rm R}$, $\Omega_{\rm M}$, and $\Omega_{\rm DE}=1-\Omega_{\rm M}-\Omega_{\rm R}$ are
the radiation, matter, and DE density parameter at the present time and the function $f(a)$ is as follows,
\begin{eqnarray}
f(a)=\frac{-3}{\ln a}\int_0^{\ln a}[1+w(a')]d \ln a'.
 \end{eqnarray}
However, in the present work, we are interested in the linear growth factor taking the DM-DE interaction into account.
In the case of interacting dark sectors, the function $E(z)$ has the following form \citep{1705.09278},
\begin{eqnarray}\label{Heath1I}
E_{\rm I}(z)=[\Omega_{\rm R} (1+z)^{4}+\Omega_{\rm B} (1+z)^{3}+ \Omega_{\rm DM} (1+z)^{3-\eta}+\frac{(1+z)^{3}}{g(z)}I(z)]^{1/2}.
 \end{eqnarray}
Here, $\Omega_{\rm B}$ and $\Omega_{\rm DM}=\Omega_{\rm M}-\Omega_{\rm B}$ are the baryonic matter and DM density parameters at the present time. Moreover,
\begin{eqnarray}
g(z)=\exp(-3\int\frac{w_{\rm DE}(z)}{1+z}dz),
 \end{eqnarray}
and
\begin{eqnarray}
I(z)=\Omega_{\rm DE} g(0)+\eta \Omega_{\rm DM} \int_0^z (1+z)^{-1-\eta}g(z) dz.
 \end{eqnarray}
Fig. \ref{fig5} presents the linear growth factor for the cases of noninteracting and interacting dark sectors.
The values of different parameters in the linear growth factor are $\Omega_{\rm R}=8.6\times10^{-5}$ and $\Omega_{\rm M}=0.276$ for the noninteracting case and $\Omega_{\rm R}=8.6\times10^{-5}$, $\Omega_{\rm B}h^2=0.02240$, $\Omega_{\rm DM}h^2=0.1259$, $\eta=0.00378$, and $h=0.704$ \citep{1705.09278} for the interacting one.
The linear growth factor decreases with the redshift for both noninteracting and interacting
dark sectors. The DM-DE interaction enhances the linear growth factor. This shows that the energy transferring from DE to DM increases the growth of matter density perturbation. The effects of the DM-DE interaction on $D(z)$ are more significant at lower values of the redshift.

\section{Probability density distribution for the ellipticity of cosmic voids}

The probability density distribution for the ellipticity characterizes the shape of cosmic voids. This distribution for the void ellipticity, $\varepsilon$, at redshift $z$ is as follows \citep{1-0704.088,426-1-440},
\begin{eqnarray}\label{prob}
p(1-\varepsilon;z)&=&p(\nu;z,R_{\rm L})=\int_{\nu}^1p[\mu,\nu|\delta=\delta_{\upsilon};\sigma(z,R_{\rm L})]d\mu\nonumber \\&=&\int_{\nu}^1d\mu\frac{3375\sqrt{2}}{\sqrt{10\pi}\sigma^5(z,R_{\rm L})}\exp[-\frac{5\delta_{\upsilon}^2}{2\sigma^2(z,R_{\rm L})}
+\frac{15\delta_{\upsilon}(\lambda_1+\lambda_2)}{2\sigma^2(z,R_{\rm L})}] \exp[-\frac{15(\lambda_1^2+\lambda_1\lambda_2+\lambda_2^2)}{2\sigma^2(z,R_{\rm L})}]\nonumber \\&\times&(2\lambda_1+\lambda_2-\delta_{\upsilon})
 (\lambda_1-\lambda_2)(\lambda_1+2\lambda_2-\delta_{\upsilon})\frac{4(\delta_{\upsilon}-3)^2\mu\nu}{(\mu^2+\nu^2+1)^3},
 \end{eqnarray}
Here, $\mu$ and $\nu$ are the void's oblateness and sphericity, respectively. $R_{\rm L}$ denotes the Lagrangian void scale and $\delta_{\upsilon}$ is the density contrast threshold for the formation of a void. Moreover, $\lambda_1$ and $\lambda_2$ are the eigenvalues of the local tidal shear tensor \citep{426-1-440,Rezaei}. $\sigma(z,R_{\rm L})$ which is called the linear rms fluctuation
of the matter density field smoothed on a Lagrangian void scale of $R_{\rm L}$ at redshift $z$, is given by \citep{1-0704.088,426-1-440,ChongchitnanSilk2010}
\begin{eqnarray}\label{sig}
\sigma^2(z,R_{\rm L})\equiv D^2(z)\int_0^{\infty}\frac{k^2}{2 \pi^2}P(k)W^2(kR_{\rm L})dk.
 \end{eqnarray}
Here, the linear growth of density perturbation, $D(z)$, has been introduced in Eqs. (\ref{Heath1}), (\ref{Heath1NI}), and (\ref{Heath1I}) for noninteracting and interacting dark sectors. $W(kR_{\rm L})$, the spherical top-hat function of radius $R_{\rm L}$, has the following form,
\begin{eqnarray}
W(kR_{\rm L})=3[\frac{\sin(kR_{\rm L})}{(kR_{\rm L})^3}-\frac{\cos(kR_{\rm L})}{(kR_{\rm L})^2}].
 \end{eqnarray} Besides, $P(k)$ is the linear matter power spectrum today and depends on the matter transfer function $T(x)$  \citep{Weinberg2008,ChongchitnanSilk2010}. The standard linear matter power spectrum considering is
 \begin{eqnarray}
P(k)=AkT^2(x),
 \end{eqnarray}
In the above equation, the coefficient $A$ should be calculated with the condition $\sigma(R_{\rm L}=8h^{-1}\ {\rm Mpc})=\sigma_8$.
The cosmological parameters related to the noninteracting and interacting cases are considered to calculate the linear matter power spectrum in these two models. Besides, the values of $\sigma_8$ and $H_0$ which have been applied in this work are $\sigma_8=0.776$ and $H_0=100h\ km\ s^{-1}\ {\rm Mpc}^{-1}$ for the noninteracting case and $\sigma_8=0.840$ and $H_0=68.13\ km\ s^{-1}\ {\rm Mpc}^{-1}$ for the interacting one \citep{1705.09278}.

\begin{figure*}
\vspace*{1cm}       % Give the correct figure height in cm
\includegraphics[width=0.30\textwidth]{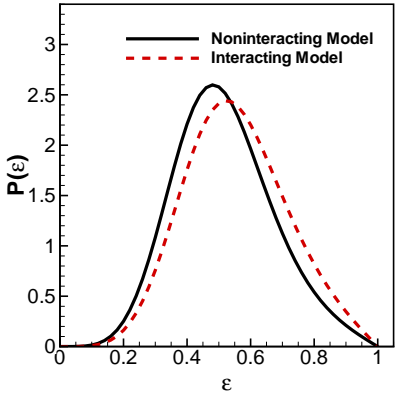}
\caption{Probability density distribution, $P(\varepsilon)$, in noninteracting and interacting models. The values of the redshift and void scale have been considered as $z=1.0$ and $R_{\rm L}=4 h^{-1}{\rm Mpc}$.}
\label{fig4}
\end{figure*}

The probability density distribution for the ellipticity of cosmic voids considering noninteracting and interacting models has been plotted in Fig. \ref{fig4}. We have set the void density contrast
$\delta_{\upsilon}=-0.9$. The probability density distribution in interacting model is shifted toward the higher values of the ellipticity compared to the noninteracting one. Thus, when the energy transfers from DE to DM, the cosmic voids are less spherical. The DE-DM coupling leads to higher values of the ellipticity of cosmic voids.

\section{{RESULTS}}

{In this section, we report our results regarding the shape of cosmic voids in the interacting cosmological model. The dependency of ellipticity of cosmic voids on the redshift as well as the void scale are explained. We also compare the results in the interacting model with those in $\Lambda$CDM one.}

\subsection{Mean ellipticity and maximum ellipticity of cosmic voids: {Redshift dependency}}

The mean ellipticity of cosmic voids which can be obtained using the probability density distribution \citep{1-0704.088},
\begin{eqnarray}
<\varepsilon>=\int_0^1\varepsilon p(\varepsilon;R_{\rm L},z) d\varepsilon,
 \end{eqnarray}
depends on the redshift, void scale, and the coupling between dark sectors.
Moreover, the probability density distribution has a maximum value at a certain value of the ellipticity called maximum ellipticity presented by $\varepsilon_{\rm max}$ which is also affected by the redshift, void scale, and DE-DM coupling.
\begin{figure*}
\vspace*{1cm}       % Give the correct figure height in cm
\includegraphics[width=0.6\textwidth]{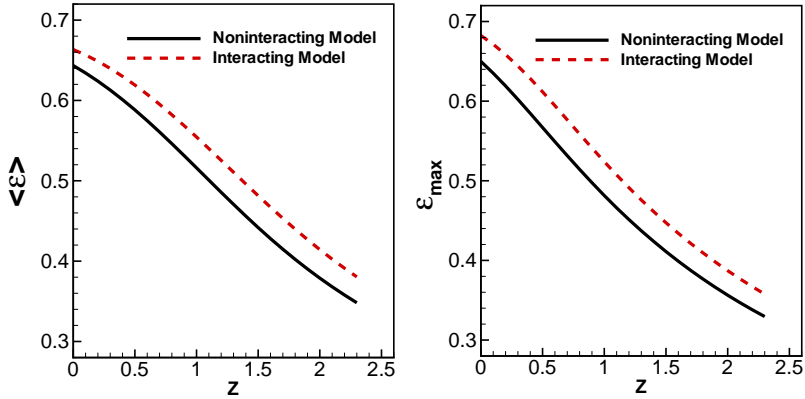}
\caption{Left: mean ellipticity, $<\varepsilon>$, and Right: maximum ellipticity, $\varepsilon_{\rm max}$, of cosmic voids in noninteracting and interacting models versus the redshift, $z$. The value of the void scale has been considered as $R_{\rm L}=4 h^{-1}{\rm Mpc}$.}
\label{fig6}
\end{figure*}
Fig. \ref{fig6} presents the values of $<\varepsilon>$ and $\varepsilon_{\rm max}$ versus the redshift.
{Both $<\varepsilon>$ and $\varepsilon_{\rm max}$ reduce as the redshift increases in agreement with the results of N-body simulations of structure formation in dynamical DE cosmologies \citep{426-1-440}.}
Considering the interacting model, $<\varepsilon>$ and $\varepsilon_{\rm max}$ show higher values at each redshift.
{In fact, the mean ellipticity of cosmic voids in the interacting model is affected by the level of clustering, i.e. $\sigma_8$.
This result has also been confirmed by N-body simulations in dynamical DE cosmologies \citep{426-1-440}.
A strong correlation between $\sigma_8$ and the mean ellipticity has been predicted in N-body simulations of structure formation \citep{426-1-440}.}
\begin{figure*}
\vspace*{1cm}       % Give the correct figure height in cm
\includegraphics[width=.6\textwidth]{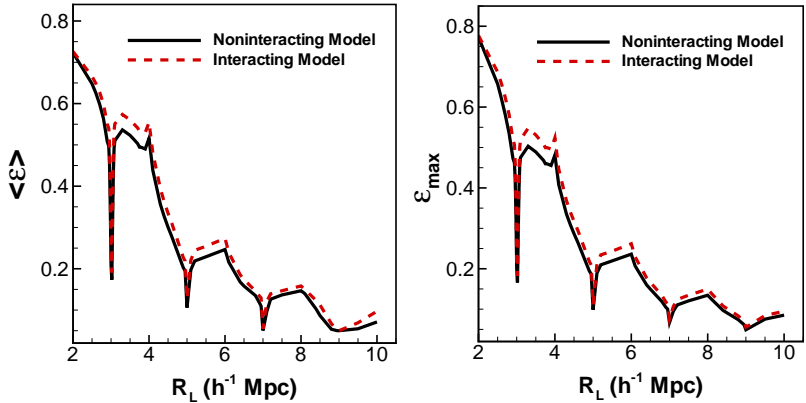}
\caption{Left: mean ellipticity, $<\varepsilon>$, and Right: maximum ellipticity, $\varepsilon_{\rm max}$, of cosmic voids in noninteracting and interacting models versus the void scale, $R_{\rm L}$. The value of the redshift has been considered as $z=1.0$.}
\label{fig8}
\end{figure*}

\subsection{Mean ellipticity and maximum ellipticity of cosmic voids: Scale dependency}

Fig. \ref{fig8} indicates that $<\varepsilon>$ and $\varepsilon_{\rm max}$ depend on the Lagrangian void scale, $R_{\rm L}$. Considering both noninteracting and interacting models,
$<\varepsilon>$ and $\varepsilon_{\rm max}$ have an oscillatory behavior with the variation of the
void scale.This behavior is along an overall reduction of $<\varepsilon>$ and $\varepsilon_{\rm max}$
with the higher values of the void scale. Our results verify that at each value of $R_{\rm L}$, DM-DE interaction increases the values of $<\varepsilon>$ and $\varepsilon_{\rm max}$. {According to high-resolution hydrodynamical N-body simulations of coupled
DE cosmologies, cosmic voids with coupled DE are also emptier and contain less neutral
hydrogen compared to $\Lambda CDM$ one \citep{1007.3736} which this effect alleviate the
tensions between simulations and observations in voids.}

%%%%%%%%%%%%%%%%%%%%%%%%%%%%%%%%%%%%%%%%%%%%%%%%%%%%%%%%%%%%%%%%%%%%%%%%%%%%%%%%%%%%%%%%%%

\section{Summary and Conclusions} \label{sec:highlight}

Evolving dark energy model constrained by the observational data has been employed to explore the ellipticity of cosmic voids in the presence of dark matter (DM) and dark energy (DE) interaction.
This interacting model based on the observational data (cosmic chronometer data, latest estimation of the local Hubble parameter value, joint light curves sample, baryon acoustic oscillation (BAO) distance measurement data set, and the cosmic microwave background (CMB) data from Planck 2015 measurements) leads to higher values of the linear growth factor compared to noninteracting one.
DM-DE interaction shifts the probability density distribution toward the greater values of ellipticity.
Our results confirm that the mean and maximum ellipticity of cosmic voids in the interacting model are larger compared to the noninteracting one.

%% If you wish to include an acknowledgments section in your paper,
%% separate it off from the body of the text using the \acknowledgments
%% command.
\acknowledgments
The author wishes to thank the Shiraz University Research Council.

%% To help institutions obtain information on the effectiveness of their
%% telescopes the AAS Journals has created a group of keywords for telescope
%% facilities.

%% Following the acknowledgments section, use the following syntax and the
%% \facility{} macro to list the keywords of facilities used in the research
%% for the paper.  Each keyword is check against the master list during
%% copy editing.  Individual instruments can be provided in parentheses,
%% after the keyword, but they are not verified.

%% This command is needed to show the entire author+affilation list when
%% the collaboration and author truncation commands are used.  It has to
%% go at the end of the manuscript.
\allauthors

%% Include this line if you are using the \added, \replaced, \deleted
%% commands to see a summary list of all changes at the end of the article.
\listofchanges


\begin{thebibliography}{}
\bibitem[Abdalla et al. (2009)]{0710.1198} Abdalla E., Abramo L. R., Sodre L., Wang B., 2009, Phys. Lett. B, {673}, 107
\bibitem[Abdalla et al. (2010)]{0910.5236} Abdalla E., Abramo L. R., Souza J. C. C. de, 2010,  Phys. Rev. D, {82}, 023508
\bibitem[Adermann et al. (2017)]{1703.04885} Adermann E., Elahi P. J., Lewis G. F., Power C., 2017,  MNRAS, {468}, 3381
\bibitem[Adermann et al. (2018)]{1807.02938} Adermann E., Elahi P. J., Lewis G. F., Power C., 2018, MNRAS, {479}, 4861
\bibitem[Amendola et al. (2007)]{0610806} Amendola L., Campos G. C., Rosenfeld R., 2007, Phys. Rev. D, {75}, 083506
\bibitem[An et al. (2018)]{1711.06799} An R., Feng C., Wang B., 2018, J. Cosmol. Astropart. Phys., {02}, 038
\bibitem[Baldi et al. (2010)]{0812.3901} Baldi M., Pettorino V., Robbers G., Springel V., 2010, MNRAS, {403}, 1684
\bibitem[Baldi \& Viel (2010)]{1007.3736} Baldi M., Viel M., 2010, MNRAS, {409}, L89
\bibitem[Barboza et al. (2009)]{Barbozaetal} Barboza E. M., Alcaniz J. S., Zhu Z. -H., Silva R., 2009, Phys. Rev. D, {80}, 043521
\bibitem[Barreiro et al. (2010)]{1004.4562} Barreiro T., Bertolami O., Torres P., 2010, MNRAS, {409}, 750
\bibitem[Bean et al. (2008)]{0709.1128} Bean R., Flanagan E. E., Trodden M., 2008, Phys. Rev. D, {78}, 023009
\bibitem[Binder \& Kremer (2006)]{Binder(2006)}  Binder J. B., Kremer G. M., 2006, Gen. Relativ. Gravit., {38}, 857
\bibitem[Biswas et al. (2010)]{1002.0014} Biswas R., Alizadeh E., Wandelt B. D., 2010, Phys. Rev. D, {82}, 023002
\bibitem[Bonnor (1957)]{Bonnor} Bonnor W. B., 1957, MNRAS, {117}, 104
\bibitem[Bonometto et al. (2015)]{1503.07875} Bonometto S. A., Mainini R., Maccio A. V., 2015, MNRAS, {453}, 1002
\bibitem[Bos et al. (2012)]{426-1-440} Bos E. G. P., Weygaert R. van de, Dolag K., Pettorino V., 2012, MNRAS, {426}, 440
\bibitem[Bruck \& Mifsud (2018)]{vandebruck2018} Bruck C. van de, Mifsud J., 2018, Phys. Rev. D, {97}, 023506
\bibitem[Cai \& Wang (2005)]{0411025} Cai R. -G., Wang A., 2005, J. Cosmol. Astropart. Phys., {03}, 002
\bibitem[Caldera-Cabral et al. (2009)]{Caldera-CabralG(2009)} Caldera-Cabral G., Maartens R., Schaefer B. M., 2009, J. Cosmol. Astropart. Phys., {07}, 027
\bibitem[Carbone et al. (2013)]{1305.0829} Carbone C., Baldi M., Pettorino V., Baccigalupi C., 2013, J. Cosmol. Astropart. Phys., {09}, 004
\bibitem[Cardenas et al. (2019)]{1812.03540} Cardenas V. H., Grandon D., Lepe S., 2019, Eur. Phys. J. C, {79}, 357
\bibitem[Carlesi et al. (2014)]{web} Carlesi E., Knebe A., Lewis G. F., Wales S., Yepes G., 2014, MNRAS, {439}, 2943
\bibitem[Chongchitnan \& Silk (2010)]{ChongchitnanSilk2010} Chongchitnan S., Silk J., 2010, Astrophys. J., {724}, 285
\bibitem[Costa et al. (2014)]{1311.7380} Costa A. A., Xu X. -D., Wang B., Ferreira E. G. M., Abdalla E., 2014, Phys. Rev. D, {89}, 103531
\bibitem[Duniya et al. (2015)]{1502.06424} Duniya D. G. A., Bertacca D., Maartens R., 2015, Phys. Rev. D, {91}, 063530
\bibitem[Dutta \& Maor (2007)]{0612027} Dutta S., Maor I., 2007, Phys. Rev. D, {75}, 063507
\bibitem[Duttaa et al. (2018)]{1707.09246} Duttaa J., Khyllepc W., Tamaninie N., 2018, J. Cosmol. Astropart. Phys., {01}, 038
{\bibitem[Elyiv et al. (2015)]{eliyv} Elyiv A., Marulli F., Pollina G., Baldi M., Branchini E., Cimatti A., Moscardini L., 2015, MNRAS, {448}, 642
}\bibitem[Fay (2016)]{1605.01644} Fay S., 2016, MNRAS, {460}, 1863
{\bibitem[Giocoli et al. (2015)]{2015Giocoli} Giocoli C., Metcalf R. B., Baldi M., Meneghetti M., Moscardini L., Petkova M., 2015, MNRAS, {452}, 2757
}\bibitem[Gonzalez \& Quiros (2008)]{0707.2089} Gonzalez T., Quiros I., 2008, Class. Quant. Grav., {25}, 175019
\bibitem[Guo et al. (2007)]{Guo(2007)} Guo Z. K., Ohta N., Tsujikawa S., 2007, Phys. Rev. D, {76}, 023508
{\bibitem[Hashim et al. (2018)]{2018hashim} Hashim M., Giocoli C., Baldi M., Bertacca D., Maartens R., 2018, MNRAS, {481}, 2933
}\bibitem[He \& Wang (2008)]{0801.4233} He J. -H., Wang B., 2008, J. Cosmol. Astropart. Phys., {06}, 010
\bibitem[He et al. (2011)]{1012.3904} He J. -H., Wang B., Abdalla E., 2011, Phys. Rev. D, {83}, 063515
\bibitem[Heath (1977)]{Heath} Heath D. J., 1977, MNRAS, {179}, 351
\bibitem[Jackson et al. (2009)]{0901.3272} Jackson B. M., Taylor A., Berera A., 2009, Phys. Rev. D, {79}, 043526
\bibitem[Koivisto \& Nunes (2013)]{1212.2541} Koivisto T. S., Nunes N. J., 2013, Phys. Rev. D, {88}, 123512
\bibitem[Kumar \& Nunes (2016)]{1608.02454} Kumar S., Nunes R. C., 2016, Phys. Rev. D, {94}, 123511
\bibitem[Kumar \& Nunes (2017)]{kumar2017} Kumar S., Nunes R. C., 2017, Phys. Rev. D, {96}, 103511
\bibitem[Lavallaz \& Fairbairn (2011)]{5-1106.1611} Lavallaz A. de , Fairbairn M., 2011, Phys. Rev. D, {84}, 083005
\bibitem[Lee \& Park (2009)]{1-0704.088} Lee J., Park D., 2009, Astrophys. J., {696}, L10
\bibitem[L'Huillier et al. (2017)]{1703.07357} L'Huillier B., Winther H. A., Mota D. F., Park C., Kim J., 2017, MNRAS, {468}, 3174
\bibitem[Linton et al. (2018)]{1711.05196} Linton M. S., Pourtsidou A., Crittenden R., Maartens R., 2018, J. Cosmol. Astropart. Phys., {04}, 043
\bibitem[Mainini (2009)]{0903.0574} Mainini R., 2009, J. Cosmol. Astropart. Phys., {04}, 017
\bibitem[Marcondes et al. (2016)]{1605.05264} Marcondes R. J. F., Landim R. C. G., Costa A. A., Wang B., Abdalla E., 2016, J. Cosmol. Astropart. Phys., {12}, 009
\bibitem[Micheletti et al. (2009)]{0902.0318} Micheletti S., Abdalla E., Wang B., 2009, Phys. Rev. D, {79}, 123506
\bibitem[Mifsud \& Bruck (2017)]{1707.07667} Mifsud J., Bruck C. van de, 2017, J. Cosmol. Astropart. Phys., {11}, 001
\bibitem[Mota et al. (2008)]{0709.2227} Mota D. F., Shaw D. J., Silk J., 2008, Astrophys. J., {675}, 29
\bibitem[Murgia et al. (2016)]{Murgia(2016)} Murgia R., Gariazzo S., Fornengo N., 2016, J. Cosmol. Astropart. Phys., {04}, 014
\bibitem[Pace et al. (2015)]{1407.7548} Pace F., Baldi M., Moscardini L., Bacon D., Crittenden R., 2015, MNRAS, {447}, 858
\bibitem[Pavon et al. (2004)]{Pavon(2004)} Pavon D., Sen S., Zimdahl W., 2004, J. Cosmol. Astropart. Phys., {05}, 009
\bibitem[Percival (2005)]{Percival} Percival W. J., 2005, A and A, {443}, 819
\bibitem[Piloyan et al. (2014)]{1401.2656} Piloyan A., Marra V., Baldi M., Amendola L., 2014, J. Cosmol. Astropart. Phys., {02}, 045
\bibitem[Pisani et al. (2015)]{1503.07690} Pisani A., Sutter P. M., Hamaus N., Alizadeh E., Biswas R., Wandelt B. D., Hirata C. M., 2015, Phys. Rev. D, {92}, 083531
\bibitem[Pollina et al. (2016)]{11} Pollina G., Baldi M., Marulli F., Moscardini L., 2016, MNRAS, {455}, 3075
\bibitem[Pourtsidou \& Tram (2016)]{1604.04222} Pourtsidou A., Tram T., 2016, Phys. Rev. D, {94}, 043518
\bibitem[Rezaei (2019)]{Rezaei} Rezaei Z., 2019, MNRAS, {487}, 2614
\bibitem[Schaefer (2008)]{0803.2239} Schaefer B. M., 2008, MNRAS, {388}, 1403
\bibitem[Simpson (2010)]{Simpson(2010)} Simpson F., 2010, Phys. Rev. D, {82}, 083505
\bibitem[Sun \& Yue (2012)]{1009.1214} Sun C. -Y., Yue R. -H., 2012, Phys. Rev. D, {85}, 043010
\bibitem[Sutter et al. (2015)]{1406.0511} Sutter P. M., Carlesi E., Wandelt B. D., Knebe A., 2015, MNRAS, {446}, L1
\bibitem[Tamanini (2015)]{1504.07397} Tamanini N., 2015, Phys. Rev. D, {92}, 043524
\bibitem[Valiviita et al. (2008)]{Valiviita (2008)} Valiviita J., Majerotto E., Maartens R., 2008, J. Cosmol. Astropart. Phys., {07}, 020
\bibitem[Valiviita \& Palmgren (2015)]{Valiviita(2015)} Valiviita J., Palmgren E., 2015, J. Cosmol. Astropart. Phys., {07}, 015
\bibitem[Wang \& Wang (2014)]{1403.4318} Wang J. S., Wang F. Y., 2014, A and A, {564}, A137
\bibitem[Wang et al. (2016)]{1603.08299} Wang B., Abdalla E., Atrio-Barandela F., Pavon D., 2016, Rep. Prog. Phys., {79}, 096901
\bibitem[Weinberg (2008)]{Weinberg2008} Weinberg S., 2008, Cosmology. Oxford Univ. Press, Oxford
\bibitem[Yang \& Xu (2014)]{1401.5177} Yang W., Xu L., 2014, J. Cosmol. Astropart. Phys., {08}, 034
\bibitem[Yang et al. (2015)]{1505.04443} Yang T., Guo Z. -K., Cai R. -G., 2015, Phys. Rev. D, {91}, 123533
\bibitem[Yang et al. (2017)]{1705.09278} Yang W., Banerjee N., Pan S., 2017, Phys. Rev. D, {95}, 123527
\bibitem[Yang et al. (2019)]{yang2018} Yang W., Pan S., Paliathanasis A., 2019, MNRAS, {482}, 1007
\bibitem[Zimdahl et al. (2001)]{Zimdahl(2001)} Zimdahl W., Pavon D., Chimento L. P., 2001, Phys. Lett. B, {521}, 133


\end{thebibliography}
\end{document}